# A Ferroic Berry Curvature Dipole in a Topological Crystalline Insulator at Room Temperature

Taiki Nishijima[1], Takuto Watanabe[3], Hiroaki Sekiguchi[3], Yuichiro Ando[1,2], Ei Shigematsu[1], Ryo Ohshima[1], Shinji Kuroda[3] and Masashi Shiraishi[1,*]

1. Department of Electronic Science and Engineering, Kyoto University, Kyoto, Kyoto 615-8510, Japan.

2. PRESTO, Japan Science and Technology Agency, Honcho, Kawaguchi, Saitama 332-0012, Japan

3. Institute of Materials Science, University of Tsukuba, 1-1-1 Tennoudai, Tsukuba, Ibaraki 305-8573, Japan

Berry curvature dipole, nonlinear Hall effect, topological crystalline insulator, ferroic switching

Abstract

The physics related to Berry curvature is now a central research topic in condensed matter physics. The Berry curvature dipole (BCD) is a significant and intriguing condensed matter phenomenon that involves inversion symmetry breaking. However, the creation and controllability of BCDs have so far been limited to far below room temperature (RT) and non-volatile (i.e., ferroic) BCDs have not yet been discovered, hindering further progress in topological physics. In this work, we demonstrate a switchable and non-volatile BCD effect at RT in a topological crystalline insulator, $Pb_{1-x}Sn_xTe$ (PST), which is attributed to ferroic distortion. Surprisingly, the magnitude of the ferroic BCD is several orders of magnitude greater than that of the non-ferroic BCDs that appear, for example, in transition metal dichalcogenides. The discovery of this ferroic and extraordinarily large BCD in PST could pave the way for further progress in topological material science and the engineering of novel topological devices.

Berry curvature plays a central role in modern condensed matter physics. It is defined in reciprocal space and provides the effect of a fictitious but emergent magnetic field in real space. As a consequence, non-zero Berry curvature endows a number of condensed matter materials with novel topological phenomena that could lead to a new era of quantum topological material science focusing on areas including topological insulators (TIs), topological crystalline insulators (TCIs), and Weyl semimetals. In particular, the Berry curvature dipole (BCD), which is the first-order moment of Berry curvature, has attracted a great deal of attention because of its production of nonlinear transport. A good example of such nonlinear transport is the nonlinear Hall effect (NLHE), the second-order Hall response, which appears in material systems with inversion symmetry breaking, such as $T_d$ phase transition metal dichalcogenides (TMDs)[1–4], strained TMD[5], Moiré systems[6,7], and monolayer ferroelectric materials[8,9]. The BCD along an arbitrary in-plane $\alpha$ direction, $D_\alpha$, in a two-dimensional system is expressed as follows:

$$D_\alpha = \int \frac{d^2\boldsymbol{k}}{(2\pi)^2} \frac{\partial f_0}{\partial k_\alpha} \Omega(\boldsymbol{k}), \tag{1}$$

where $\boldsymbol{k}$ is a wave vector, $k_\alpha$ is a wave vector along $\alpha$, $f_0$ is the Fermi distribution function, and $\Omega(\boldsymbol{k})$ is the Berry curvature. Under the time-reversal symmetric condition, $D_\alpha$ is zero in centrosymmetric materials. Meanwhile, $D_\alpha$ is nonzero when the system has broken inversion symmetry along the $\beta$ direction ($\alpha \perp \beta$), where the breaking is attributed to a crystalline structure, distortion, or displacement of specific constituent atoms. The direction and amplitude of $D_\alpha$ can be monitored via the NLHE[10], and in fact, reasonably large values of $D_\alpha$ have been reported[6]. However, there are two complications that hamper further progress in BCD physics. One is that detection of the NLHE due to a BCD is limited at low temperatures. This is mainly because a small BCD invokes a weak NLHE, because the NLHE is a second-order effect. The other is the difficulty involved in the non-volatile modulation of a BCD. Since a BCD is a built-in property in conventional materials, only external fields such as from an electrical gating or mechanical strain can tune and switch a BCD, but a BCD cannot otherwise keep its state statically.

TCIs are a class of topological quantum materials whose topological surface states are protected by mirror symmetry, unlike TIs, in which time-reversal symmetry protects their states. The topological surface state in a TCI allows an abundant physical nature as in TIs, and this nature can be controlled by mechanical strain, structural distortion, impurity doping, and temperature[11–13]. However, progress in the quest for the above-mentioned intriguing physical nature in TCIs has lagged behind that of TIs, and indeed, pioneering research on TCIs has been mainly limited to photoelectron spectroscopy[14–21] and scanning tunnelling microscopy[21–23]. The rock-salt structure IV-VI materials, $Pb_{1-x}Sn_xTe$ (PST), are a family of TCIs[12,16,24] (see Fig. 1a) and host massive or massless four Dirac cones on their (001) surfaces that reflect ferroelectric displacements of atoms at the surfaces[15,22,23]. In line with the physics of the BCD, it is envisaged that the ferroelectric structural transition of PST gives rise to structural distortion that renders the BCD tuneable and switchable for the following reason. Under the ferroelectric distortion, the mirror symmetry in PST breaks, resulting in opening gaps of the Dirac cones located perpendicular to the direction of ferroelectric distortion (Figs. 1b and 1c). Dirac fermions are massive by the gap opening and acquire finite Berry curvature (Fig. 1d). Because the Berry curvatures of the massive Dirac fermions at each of the gapped Dirac states have opposite signs due to time-reversal symmetry, the BCD arises perpendicularly to the ferroelectric distortion, yielding a nonzero NLHE[10]. More importantly, the magnitude and polarity of the BCD, which are dependent on the extent of the band-gap opening and the concomitant generation of massive Dirac fermions by the ferroelectric distortion of PST, are controllable if the ferroelectric distortion is switched by an external electric field, and furthermore, the BCD in PST is non-volatile because the extent of the ferroelectric distortion governs the creation of the BCD. Hence, utilization of the ferroelectric nature of PST allows for the pioneering of a novel and undiscovered physical phenomenon—a tuneable, switchable, and non-volatile BCD, specifically a ferroic BCD, detected via modulation of the NLHE by an external electric field. In addition, the discovery of this novel effect enables the acceleration of studies on TCIs, for which research progress has been slower than for other topological quantum materials.

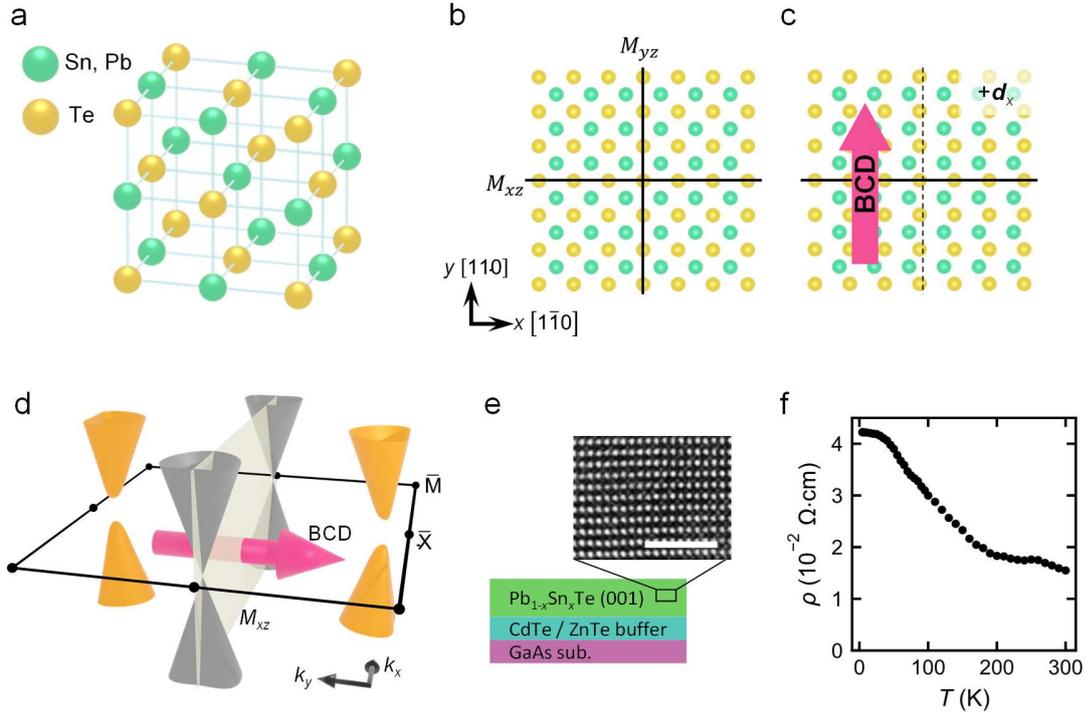

**Figure 1** Crystal structure, ferroelectric distortion and sample structure of Pb$_{1-x}$Sn$_x$Te (PST). (a) The rock-salt crystal structure of PST. Sn, Pb cations and Te anions are coloured green and yellow, respectively. (b) The undistorted (001) surface of the PST where two mirror planes denoted as $M_{xz}$ and $M_{yz}$ are both protected (indicated by black solid lines). In this paper, the $x$ and $y$ axes are defined in the [1$\bar{1}$0] and [110] lattice orientation, respectively. (c) The distorted (001) surface of the PST, where the ferroelectric displacement of atoms takes place along the $x$ direction. Owing to the distortion, a mirror plane $M_{xz}$ (indicated by a black dashed line) is broken and a finite BCD is present. (d) A schematic of the surface band structure in the first Brillouin zone when the system is under ferroelectric distortion along the $x$ direction. PST has four Dirac cones near each $\bar{X}$ point. Breaking of one mirror plane (here, $M_{yz}$ is broken) opens gaps only in two Dirac cones along $k_y$ (coloured orange) while the other two (coloured grey) remain gapless. The gapped Dirac cones gain finite Berry curvature with opposite signs to one another due to time-reversal symmetry. As a result of their segregation in momentum space, a finite BCD arises along $k_y$. (e) A sample stack structure and {110} cross-sectional TEM image of the PST. The white scale bar in the TEM image indicates 2 nm. The PST (001) layer is epitaxially grown on a CdTe / ZnTe buffer layer on a GaAs substrate. (f) The temperature dependence of the resistivity of the PST, showing semiconductive behaviour in which the resistivity increases with decreasing temperature.

An Sb-doped PST single crystal was grown on insulative CdTe/ZnTe buffer layers on a GaAs(001) substrate by molecular beam epitaxy. Energy dispersive x-ray spectrometry (EDX) from tunnelling electron microscopy (TEM) revealed that the composition of $x$ was 52%, i.e., $Pb_{0.48}Sn_{0.52}Te$.. From x-ray diffraction measurements, the crystal orientation was determined to be $(001)[110]_{PST} // (001)[110]_{GaAs}$ (see S.I. No. I). Figure 1f shows the temperature-dependence of the resistivity of the $Pb_{0.48}Sn_{0.52}Te$, where a typical TI-like dependence is found. Indeed, the resistivity monotonically increases with decreasing temperature down to 50 K and saturates below 50 K, and this is attributed to the fact that the bulk conduction is suppressed and carrier transport in the topological surface state is dominant. Such a semiconducting bulk conduction is realized by Sb doping that compensates for Sn vacancy sites, resulting in surface dominant transport at low temperature. A similar characteristic temperature dependence of the resistivity is widely observed in many TIs, such as $(Bi,Sb)_2(Te,Se)_3$ and $Tl_{1-x}Bi_{1+x}Se_2$[25,26]. Furthermore, the composition of the Sn was identified to be 52%, allowing for the TCI character[27]. Thus, it is concluded from the results that the PST used in this study is TCI, a topologically non-trivial material.

As mentioned previously, the ferroelectric character of PST allows observation of the NLHE. We measured the nonlinear Hall (NLH) voltage $V_{\hat{a}-\hat{b}\hat{b}}^{2\omega}$ using the a.c. lock-in technique (see Fig. 2a). $\hat{b}$ and $\hat{a}$ denote the in-plane directions, along which a sinusoidal electric current $I_{\hat{b}}^{\omega} =$ $|I_{\hat{b}}^{\omega}|\sin(\omega t)$ was applied and a second harmonic voltage measurement was implemented, respectively. The NLH voltage, which arises as the second harmonic transverse voltage along $\hat{a}$, is given by $V_{\hat{a}-\hat{b}\hat{b}}^{2\omega} \propto D_{\hat{b}} \cdot \left(I_{\hat{b}}^{\omega}\right)^2$, where $D_{\hat{b}}$ is the BCD along $\hat{b}$. The $|I_{\hat{y}}^{\omega}|$ dependence of $V_{\hat{x}-\hat{y}\hat{y}}^{2\omega}$ is shown in Fig. 2b (see also Fig. 2a for the definition of the $x$ and $y$ coordinates). A noticeable quadratic dependence on $|I_{\hat{y}}^{\omega}|$, which is a manifestation of the NLHE, is measured in the PST (the second harmonic longitudinal voltage, $V_{\hat{y}-\hat{y}\hat{y}}^{2\omega}$, is considerably smaller than $V_{\hat{x}-\hat{y}\hat{y}}^{2\omega}$, which supports the successful detection of the NLHE[2,4] (see S.I. No. IV)), and more surprisingly, the NLHE is substantial up to 300 K. The NLHE is monotonically enhanced at lower temperatures, which can be explained by the suppression of the parallel conduction of the bulk and topological surface states of the PST. Therefore, the significant enhancement of $V_{\hat{x}-\hat{y}\hat{y}}^{2\omega}$ at low temperatures demonstrates that the NLH voltage along $\hat{x}$ in the PST is attributed to the topological surface state of the PST, where the BCD is created along $\hat{y}$ due to inversion symmetry breaking along $\hat{x}$ (note that the ferroelectric lattice distortion along $\hat{x}$ creates massive Dirac fermions in a pair of Dirac cones along $k_y$, resulting in the BCD along $\hat{y}$). As a control experiment, we selected PbTe (PT), a topologically trivial material, and conducted the same measurement. As shown in Fig. 2c, the NLHE is not observed in the PT even at a low temperature (S.I. No. IV), which is compelling evidence that the topological nature of the surface state of the PST is the key to the appearance of the NLHE. From fitting results using the quadratic function $V_{\hat{x}-\hat{y}\hat{y}}^{2\omega} = \chi\left(I_{\hat{y}}^{\omega}\right)^2$, we extracted $\chi$, an index of the strength of the NLH response, for the PST and the PT, and the results are summarized in Fig. 2d. The coefficient $\chi$ of the PST at 5 K is 35 times greater than that at 300 K. As discussed above, the NLHE nature is ascribed to the topological surface state of PST. Hence, we deduce that further enhancement of the NLHE can be realized by decreasing the thickness of the PST to suppress the bulk contribution to the carrier transport.

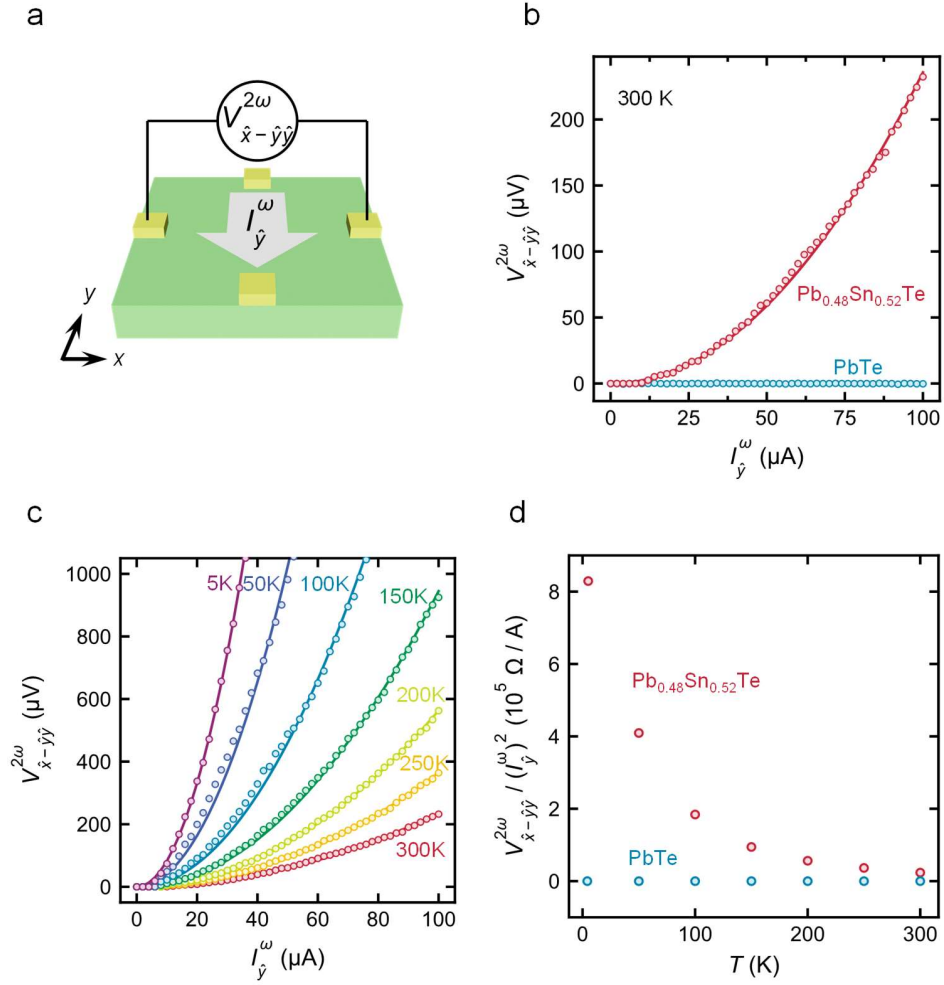

**Figure 2** Nonlinear Hall effect in PST and a comparison with PT. (a) A schematic of the experimental setup for the NLHE measurement. (b) The second harmonic transverse voltage $V_{\hat{x}-\hat{y}\hat{y}}^{2\omega}$ of the a.c. current along $y$ at 300 K. The results are coloured red and blue for PST and PT, respectively. (c) The temperature dependence of the $V_{\hat{x}-\hat{y}\hat{y}}^{2\omega} - I_{\hat{b}}^{\omega}$ characteristics in PST. In all of the measured range, $5 - 300$ K, a quadratic dependence is observed. (d) The strength of the NLHE summarized as a function of temperature. The PT does not exhibit NLHE.

To further investigate the relevance of the successful detection of the NLHE by ferroelectric distortion in the PST, we performed NLH voltage measurements after applying the pulsed electric field to control the ferroelectric distortion and its direction. A pulsed electric field of 15 V/cm, that is more than ten times greater than that of the input sinusoidal electric field of the NLH voltage measurements, was applied for 1 s along $\pm x$ and $\pm y$ ($E_{\pm x}^{\text{pulse}}$ and $E_{\pm y}^{\text{pulse}}$) at 300 K, resulting in ferroelectric lattice distortion along the $\pm x$ and $\pm y$ directions. Since (1) the BCD is created by lattice inversion symmetry breaking due to the ferroelectric distortion and (2) the direction of the ferroelectric distortion is controlled by a polarity of the pulse electric field, the BCD can be modulated by the repeated application of a pulse electric field and the polarity of the BCD is controllable. Hence, the absolute value of the NLH voltage $V_{\hat{x}-\hat{y}\hat{y}}^{2\omega}$ can increase quadratically as a function of $I_{\hat{y}}^{\omega}$ because $V_{\hat{x}-\hat{y}\hat{y}}^{2\omega} \propto D_y \cdot \left(I_{\hat{y}}^{\omega}\right)^2$, and its polarity can be changed by the $E_{\pm x}^{\text{pulse}}$. As can be seen in Fig. 3a, the NLH voltage $V_{\hat{x}-\hat{y}\hat{y}}^{2\omega}$ depends quadratically on $I_{\hat{y}}^{\omega}$ and the polarity reversal of $V_{\hat{x}-\hat{y}\hat{y}}^{2\omega}$ under the application of $E_{+x}^{\text{pulse}}$ and $E_{-x}^{\text{pulse}}$ is successfully detected. These results can be understood from the picture of ferroic BCD creation, where the lattice distortion of the PST breaks a mirror plane ($1\bar{1}0$) and creates a BCD along the $y$ direction (see Figs. 3b and 3c, and also S.I. No. II for the detail of the symmetry analyses). For comparison, we also performed the same measurements by applying the pulsed electric field along the $\pm y$ direction and confirmed that the $V_{\hat{x}-\hat{y}\hat{y}}^{2\omega}$ is negligibly small (Fig. 3d). The missing $V_{\hat{x}-\hat{y}\hat{y}}^{2\omega}$ is rationalized as follows: As shown in Figs. 3e and 3f, the BCD is created along the $\pm x$ directions under the application of the pulsed electric field along the $\pm y$ direction. Because the NLH voltage is detected along the $y$ direction, the BCD-induced NLHE that appears along the $x$ direction is not detectable in this setup. These findings sufficiently coincide with the physical picture that the creation of BCD by the ferroelectric distortion attributes to the NLHE, i.e., the directions of the distortion and the BCD are perpendicular and the NLH voltage is detected perpendicularly to the BCD.

To confirm that the ferroelectric distortion definitely produces the NLHE, we measured $V_{\hat{x}-\hat{y}\hat{y}}^{2\omega}$ with a sweeping $E_x^{\text{pulse}}$ upwards and downwards. The observation of a hysteretic loop of $V_{\hat{x}-\hat{y}\hat{y}}^{2\omega}$ is expected, because the BCD can be modulated via ferroelectric distortion. As expected, clear hysteresis of $V_{\hat{x}-\hat{y}\hat{y}}^{2\omega}$ as a function of $E_x^{\text{pulse}}$ was measured (fig. 3g). Here, $V_{\hat{x}-\hat{y}\hat{y}}^{2\omega} \propto D_y \left( I_y^{\omega} \right)^2$, so the observed $V_{\hat{x}-\hat{y}\hat{y}}^{2\omega}$ simply depends on the projection of the BCD along the $y$ direction, which is controlled by the change in the ferroelectric distortion by the $E_x^{\text{pulse}}$. The observed multilevel states remain non-volatile without an external electric field, and hence, the successful observation of the hysteresis of the NLH voltages underscores the validity of our assertion for the underlying physics, i.e., the BCD in PST is the key to the generation of the NLHE at room temperature and is electrically controllable through the ferroelectric distortion. The absence of hysteresis saturation within the measured electric field range indicates that the ferroelectric distortion forms multiple domains and that a higher electric field is required for the hysteresis saturation to occur. We also explored the hysteresis characteristics of NLHE at low temperature (see SI No. V).

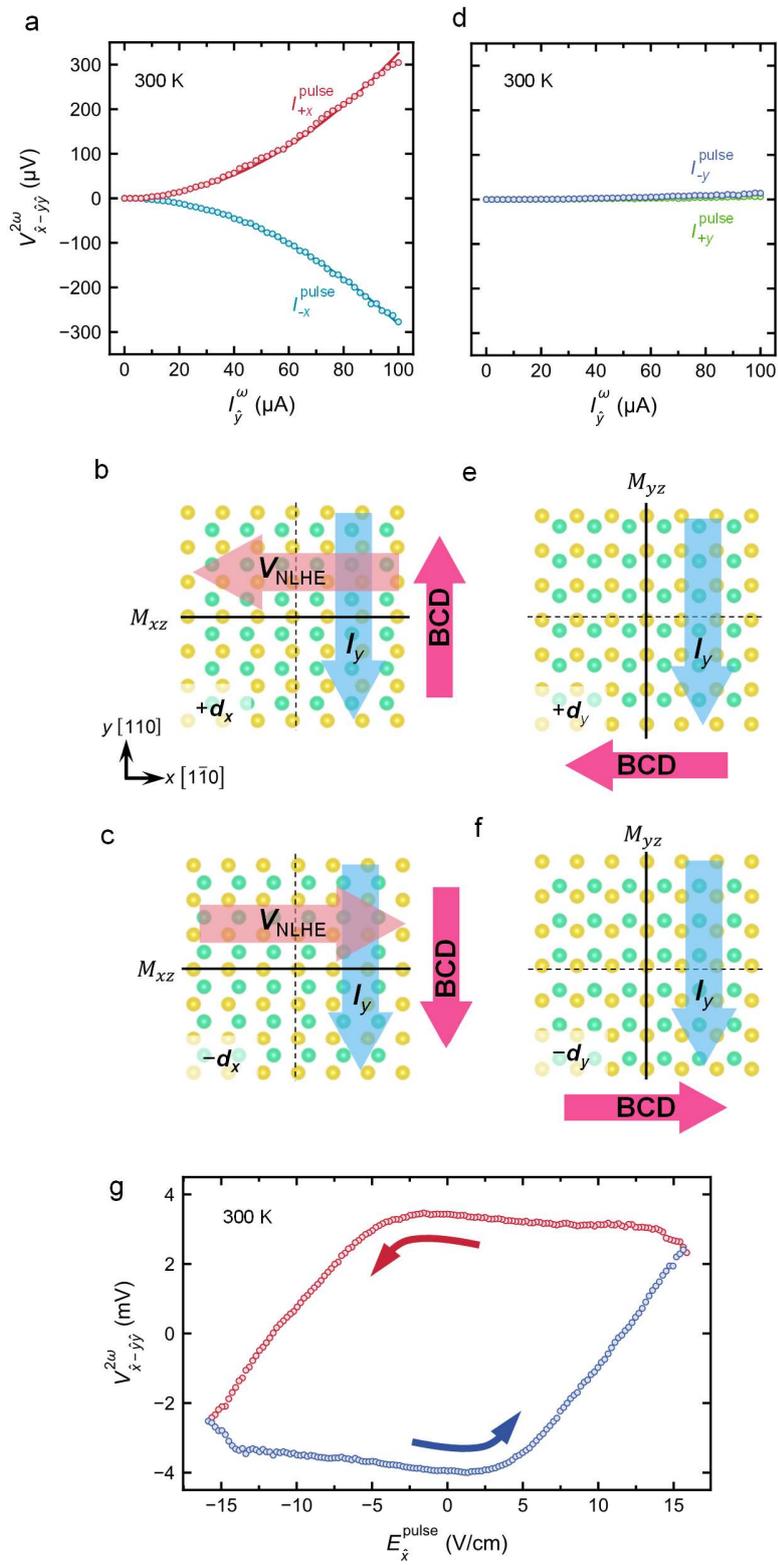

**Figure 3** Nonlinear Hall effect in PST with mirror symmetry control. (a) $V_{\hat{x}-\hat{y}\hat{y}}^{2\omega} - I_{\hat{b}}^{\omega}$ characteristics after application of a pulsed electric field of 15 V/cm from $+x$ (red) and $-x$ (cyan). The inset is a schematic of the measurement setup. (b,c) Schematics showing the correlation between directions of the ferroelectric distortion $d_x$ and BCD with the (001) surface of PST that describes the result of (a). Due to the ferroelectric distortion $+d_x$, a BCD along the $y$ direction is present (b). Under the injection of current $I_y$ parallel to the BCD, the NLHE generates a finite voltage $V_{\text{NLHE}}$ in the $x$ direction. When the ferroelectric distortion is reversed to $-d_x$, the direction of the BCD flips and the resulting $V_{\text{NLHE}}$ is generated in the opposite direction (c). (d) $V_{\hat{x}-\hat{y}\hat{y}}^{2\omega} - I_{\hat{b}}^{\omega}$ characteristics after application of a pulsed electric field of 15 V/cm from $+y$ (green) and $-y$ (blue). (e,f) Schematics when the ferroelectric distortion is in the $y$ direction in accordance with (b) and (c), respectively. Here BCD is oriented in the $x$ direction, so the injected current is orthogonal to the BCD. Therefore, the BCD-induced NLHE vanishes. (g) $V_{\hat{x}-\hat{y}\hat{y}}^{2\omega}$ measured with a sweeping pulsed electric field, exhibiting clear hysteresis behaviour and sign reversal. Red (blue) circles show the down sweep (up sweep). $I_{\hat{b}}^{\omega} = 500$ μA.

Evaluation of the magnitude of the BCD and comparison with those of other materials systems where a non-ferroic BCD allows the NLHE is significant. Since the PST has parallel conduction of the bulk and topological surface states, quantitative estimation of the magnitude of the BCD is somewhat complicated at room temperature due to the contribution of the bulk state. However, at low temperatures, the bulk conduction is suppressed and the surface transport of carriers is dominant. Therefore, we focus only on the estimation of the BCD at 5 K by postulating the complete suppression of the bulk transport. The nonlinear Hall current at low frequency is given by $J^{2\omega} = \frac{e^3\tau}{2\hbar^2}D(E^\omega)^2$ where $e$ is the elementary charge, $\tau$ is the scattering time, $\hbar$ is the Dirac constant, $D$ is the BCD, and $E^\omega$ is the electric field[2,10]. Using $\chi = \frac{e^3\tau}{2\hbar^2}\frac{1}{w\sigma^3}D$, this formula can be transformed into a fitting formula for the experimental result, $V_{a-bb}^{2\omega} = \chi(I_b^\omega)^2$, where $w$ is the width of the sample and $\sigma$ is the conductivity. At 5 K, from the experimental result shown in Fig. 2c (the ferroelectric distortion is oriented along the $x$ direction by the pulsed electric field), $\chi = V_{a-bb}^{2\omega}/(I_b^\omega)^2$ was found to be $8.29\times10^5$ V/A$^2$ in the PST. Using $\tau$ = 6 ps[28] and the conductivity measured by the van der Pauw method (Fig. 1f), the BCD of the PST was estimated to be 150 nm. It is notable that this estimation provides a lower limit because the bulk conduction is not completely suppressed in the actual experiments. This estimated BCD (150 nm) is more than one order of magnitude greater than that of non-ferroic BCD materials such as TMD (5 nm$^2$). It is also noted that the magnitude is two orders of magnitude larger than the theoretically predicted value for SnTe[10]. However, we emphasize that similar unprecedentedly large BCDs have been frequently reported in previous studies using twisted double bilayer graphene[6] and twisted bilayer WSe$_2$[6,7]. One possible origin of such discrepancies stems from extrinsic mechanisms, such as skew scattering and side-jumps[29–33]. However, it should be emphasized that the contribution of skew scattering to nonlinear transversal voltage generation reported in Refs. 32,33 can be neglected in our system because it is observed in a symmetry-protected system that enables suppression of the BCD and is considered to be independent of the switching of ferroelectric distortion. Regarding the other extrinsic contributions, although the scaling law has been proposed as a potential approach to distinguish the physical origins of the NLHE[29], there is a subtlety about whether the scaling law is obtained in the PST because the control of conductivity or disorders of PST is not easy without changing its ferroelectric nature and lattice inversion symmetry. Consequently, these extrinsic contributions, which are challenging to isolate, can be deemed as one of the factors accounting for the discrepancy between the BCD values calculated theoretically and experimentally.

As mentioned previously, the experimental results unambiguously substantiate the existence of a ferroic BCD in PST resulting in the observation of the NLHE at room temperature. In this paragraph, opportunities to create innovative electric devices using ferroic characteristics are discussed. Given that the effect can manifest itself at room temperature, a possible application is non-volatile BCD memory, which can be simply achieved using a single ferroic BCD material with a Hall bar structure (see also SI No. VI). Because the direction of the external electric field and the reading direction are orthogonal, a single strip allows for multiple-bit storage. The large BCD in PST discovered in this study has the great advantage of allowing for downsizing of the device. Furthermore, because PST is a nonmagnetic ferroelectric metal, in which an external electric field can be screened by surface carriers, the proposed BCD memory is expected to retain information firmly against disturbances such as magnetic fields and electric fields. In BCD memory, the fabrication of three-dimensionally integrated memory is also possible only by fabricating heterostructures of BCD and non-BCD materials such as PST and PT. In this system, only the topological surface state with BCD characteristics is conductive and data can be stored in two surface states in each PST layer.

To conclude, a ferroic BCD has been successfully detected at RT using a TCI, Sb-doped PST. The ferroic BCD is tuneable, switchable, and non-volatile at RT, and the magnitude of the BCD greatly surpasses that of a material with a non-ferroic nature. An interplay between the topological surface state protected by inversion symmetry and ferroic atom distortion in PST at RT renders the ferroic BCD resilient against thermal perturbations. Although the progress of research on TCIs has lagged behind that of the other topological quantum materials since the advent of TCIs, this newly discovered ferroic BCD in PST should enable the acceleration of research on topological physics targeting TCIs.

ASSOCIATED CONTENT

**Supporting Information**. The Supporting Information is available. Sample characterization, symmetry analysis on BCD, ferroic nature of PST, additional NLHE data, hysteresis of NLHE at low temperature, an application idea of ferroic BCD.

AUTHOR INFORMATION

**Corresponding Author**

* Email: shiraishi.masashi.4w@kyoto-u.ac.jp

**Author Contributions**

T.N. conceived and conducted transport experiments supervised by M.S., Y.A., R.O. and E.S. T.W., H.S. and S.K. grew the PST single crystal. T.N., T.W. and H.S. characterized the sample. T.N., Y.A. and M.S. wrote the manuscript, with input from all authors. All authors discussed the results and reviewed the manuscript.

**Notes**

The authors declare no competing financial interest.

ACKNOWLEDGMENT

A part of this study was supported by Japan Society for the Promotion of Science (KAKENHI Grants No. 22H00214, and No. 20K22413) and Japan Science and Technology Agency (JST), JST-PRESTO 'Information Carrier' (Grant No. JPMJPR20B2).

REFERENCES

(Word Style "TF_References_Section"). References are placed at the end of the manuscript. Authors are responsible for the accuracy and completeness of all references. Examples of the recommended format for the various reference types can be found at

(1)     Kang, K.; Li, T.; Sohn, E.; Shan, J.; Mak, K. F. Nonlinear Anomalous Hall Effect in Few-Layer $WTe_2$. *Nat. Mater.* **2019**, *18* (4), 324–328. https://doi.org/10.1038/s41563-019-0294-7.

(2)     Ma, Q.; Xu, S.-Y.; Shen, H.; MacNeill, D.; Fatemi, V.; Chang, T.-R.; Mier Valdivia, A. M.; Wu, S.; Du, Z.; Hsu, C.-H.; Fang, S.; Gibson, Q. D.; Watanabe, K.; Taniguchi, T.; Cava, robert J.; Kaxiras, E.; Lu, H.-Z.; Lin, H.; Fu, L.; Gedik, N.; Jarillo-Herrero, P. Observation of the Nonlinear Hall Effect under Time-Reversal-Symmetric Conditions. *Nature* **2019**, *565* (7739), 337–342. https://doi.org/10.1038/s41586-018-0807-6.

(3)     Xiao, J.; Wang, Y.; Wang, H.; Pemmaraju, C. D.; Wang, S.; Muscher, P.; Sie, E. J.; Nyby, C. M.; Devereaux, T. P.; Qian, X.; Zhang, X.; Lindenberg, A. M. Berry Curvature Memory through Electrically Driven Stacking Transitions. *Nat. Phys.* **2020**, *16* (10), 1028–1034. https://doi.org/10.1038/s41567-020-0947-0.

(4)     Kumar, D.; Hsu, C.-H.; Sharma, R.; Chang, T.-R.; Yu, P.; Wang, J.; Eda, G.; Liang, G.; Yang, H. Room-Temperature Nonlinear Hall Effect and Wireless Radiofrequency Rectification in Weyl Semimetal $TaIrTe_4$. *Nat. Nanotechnol.* **2021**, *16* (4), 421–425. https://doi.org/10.1038/s41565-020-00839-3.

(5)     Qin, M.-S.; Zhu, P.-F.; Ye, X.-G.; Xu, W.-Z.; Song, Z.-H.; Liang, J.; Liu, K.; Liao, Z.-M.
        Strain Tunable Berry Curvature Dipole, Orbital Magnetization and Nonlinear Hall Effect
        in $WSe_2$ Monolayer. *Chinese Phys. Lett.* **2021**, *38* (1), 017301.
        https://doi.org/10.1088/0256-307X/38/1/017301.

(6)     Sinha, S.; Adak, P. C.; Chakraborty, A.; Das, K.; Debnath, K.; Sangani, L. D. V.;
        Watanabe, K.; Taniguchi, T.; Waghmare, U. V.; Agarwal, A.; Deshmukh, M. M. Berry
        Curvature Dipole Senses Topological Transition in a Moiré Superlattice. *Nat. Phys.* **2022**,
        *18* (7), 765–770. https://doi.org/10.1038/s41567-022-01606-y.

(7)     Huang, M.; Wu, Z.; Hu, J.; Cai, X.; Li, E.; An, L.; Feng, X.; Ye, Z.; Lin, N.; Tuen Law,
        K.; Wang, N. Giant Nonlinear Hall Effect in Twisted Bilayer $WSe_2$. *Natl. Sci. Rev.* **2022**,
        nwac232. https://doi.org/10.1093/nsr/nwac232/6769863.

(8)     Kim, J.; Kim, K.-W.; Shin, D.; Lee, S.-H.; Sinova, J.; Park, N.; Jin, H. Prediction of
        Ferroelectricity-Driven Berry Curvature Enabling Charge- and Spin-Controllable
        Photocurrent in Tin Telluride Monolayers. *Nat. Commun.* **2019**, *10* (1), 3965.
        https://doi.org/10.1038/s41467-019-11964-6.

(9)     Jin, K.-H.; Oh, E.; Stania, R.; Liu, F.; Yeom, H. W. Enhanced Berry Curvature Dipole and
        Persistent Spin Texture in the Bi(110) Monolayer. *Nano Lett.* **2021**, *21* (22), 9468–9475.
        https://doi.org/10.1021/acs.nanolett.1c02811.

(10)    Sodemann, I.; Fu, L. Quantum Nonlinear Hall Effect Induced by Berry Curvature Dipole
        in Time-Reversal Invariant Materials. *Phys. Rev. Lett.* **2015**, *115* (21), 216806.
        https://doi.org/10.1103/PhysRevLett.115.216806.

(11)  Ando, Y.; Fu, L. Topological Crystalline Insulators and Topological Superconductors: From Concepts to Materials. *Annu. Rev. Condens. Matter Phys.* **2015**, *6* (1), 361–381. https://doi.org/10.1146/annurev-conmatphys-031214-014501.

(12)  Wang, J.; Wang, N.; Huang, H.; Duan, W. Electronic Properties of SnTe-Class Topological Crystalline Insulator Materials. *Chinese Phys. B* **2016**, *25* (11), 117313. https://doi.org/10.1088/1674-1056/25/11/117313.

(13)  Serbyn, M.; Fu, L. Symmetry Breaking and Landau Quantization in Topological Crystalline Insulators. *Phys. Rev. B* **2014**, *90* (3), 035402. https://doi.org/10.1103/PhysRevB.90.035402.

(14)  Dziawa, P.; Kowalski, B. J.; Dybko, K.; Buczko, R.; Szczerbakow, A.; Szot, M.; Łusakowska, E.; Balasubramanian, T.; Wojek, B. M.; Berntsen, M. H.; Tjernberg, O.; Story, T. Topological Crystalline Insulator States in $Pb_{1-x}Sn_xSe$. *Nat. Mater.* **2012**, *11* (12), 1023–1027. https://doi.org/10.1038/nmat3449.

(15)  Wojek, B. M.; Berntsen, M. H.; Jonsson, V.; Szczerbakow, A.; Dziawa, P.; Kowalski, B. J.; Story, T.; Tjernberg, O. Direct Observation and Temperature Control of the Surface Dirac Gap in a Topological Crystalline Insulator. *Nat. Commun.* **2015**, *6* (1), 8463. https://doi.org/10.1038/ncomms9463.

(16)  Tanaka, Y.; Ren, Z.; Sato, T.; Nakayama, K.; Souma, S.; Takahashi, T.; Segawa, K.; Ando, Y. Experimental Realization of a Topological Crystalline Insulator in SnTe. *Nat. Phys.* **2012**, *8* (11), 800–803. https://doi.org/10.1038/nphys2442.

(17)   Fu, L.; Kane, C. L. Topological Insulators with Inversion Symmetry. *Phys. Rev. B* **2007**, *76* (4), 045302. https://doi.org/10.1103/PhysRevB.76.045302.

(18)   Yan, C.; Liu, J.; Zang, Y.; Wang, J.; Wang, Z.; Wang, P.; Zhang, Z. D.; Wang, L.; Ma, X.; Ji, S.; He, K.; Fu, L.; Duan, W.; Xue, Q. K.; Chen, X. Experimental Observation of Dirac-like Surface States and Topological Phase Transition in $Pb_{1-x}Sn_xTe$ (111) Films. *Phys. Rev. Lett.* **2014**, *112* (18), 186801. https://doi.org/10.1103/PhysRevLett.112.186801.

(19)   Polley, C. M.; Dziawa, P.; Reszka, A.; Szczerbakow, A.; Minikayev, R.; Domagala, J. Z.; Safaei, S.; Kacman, P.; Buczko, R.; Adell, J.; Berntsen, M. H.; Wojek, B. M.; Tjernberg, O.; Kowalski, B. J.; Story, T.; Balasubramanian, T. Observation of Topological Crystalline Insulator Surface States on (111)-Oriented $Pb_{1-x}Sn_xSe$ Films. *Phys. Rev. B - Condens. Matter Mater. Phys.* **2014**, *89* (7), 075317. https://doi.org/10.1103/PhysRevB.89.075317.

(20)   Ito, H.; Otaki, Y.; Tomohiro, Y.; Ishida, Y.; Akiyama, R.; Kimura, A.; Shin, S.; Kuroda, S. Observation of Unoccupied States of SnTe(111) Using Pump-Probe ARPES Measurement. *Phys. Rev. Res.* **2020**, *2* (4), 043120. https://doi.org/10.1103/PhysRevResearch.2.043120.

(21)   Gyenis, A.; Drozdov, I. K.; Nadj-Perge, S.; Jeong, O. B.; Seo, J.; Pletikosić, I.; Valla, T.; Gu, G. D.; Yazdani, A. Quasiparticle Interference on the Surface of the Topological Crystalline Insulator $Pb_{1-x}Sn_xSe$. *Phys. Rev. B* **2013**, *88* (12), 125414. https://doi.org/10.1103/PhysRevB.88.125414.

(22)     Okada, Y.; Serbyn, M.; Lin, H.; Walkup, D.; Zhou, W.; Dhital, C.; Neupane, M.; Xu, S.;

Wang, Y. J.; Sankar, R.; Chou, F.; Bansil, A.; Hasan, M. Z.; Wilson, S. D.; Fu, L.;

Madhavan, V. Observation of Dirac Node Formation and Mass Acquisition in a

Topological Crystalline Insulator. *Science* **2013**, *341* (6153), 1496–1499.

https://doi.org/10.1126/science.1239451.

(23)     Zeljkovic, I.; Okada, Y.; Serbyn, M.; Sankar, R.; Walkup, D.; Zhou, W.; Liu, J.; Chang,

G.; Wang, Y. J.; Hasan, M. Z.; Chou, F.; Lin, H.; Bansil, A.; Fu, L.; Madhavan, V. Dirac

Mass Generation from Crystal Symmetry Breaking on the Surfaces of Topological

Crystalline Insulators. *Nat. Mater.* **2015**, *14* (3), 318–324.

https://doi.org/10.1038/nmat4215.

(24)     Hsieh, T. H.; Lin, H.; Liu, J.; Duan, W.; Bansil, A.; Fu, L. Topological Crystalline

Insulators in the SnTe Material Class. *Nat. Commun.* **2012**, *3* (1), 982.

https://doi.org/10.1038/ncomms1969.

(25)     Nastos, F.; Rioux, J.; Strimas-Mackey, M.; Mendoza, B. S.; Sipe, J. E. Full Band Structure

LDA and K·p Calculations of Optical Spin-Injection. *Phys. Rev. B* **2007**, *76* (20), 205113.

https://doi.org/10.1103/PhysRevB.76.205113.

(26)     Ando, Y.; Hamasaki, T.; Kurokawa, T.; Ichiba, K.; Yang, F.; Novak, M.; Sasaki, S.;

Segawa, K.; Ando, Y.; Shiraishi, M. Electrical Detection of the Spin Polarization Due to

Charge Flow in the Surface State of the Topological Insulator $Bi_{1.5}Sb_{0.5}Te_{1.7}Se_{1.3}$. *Nano

Lett.* **2014**, *14* (11), 6226–6230. https://doi.org/10.1021/nl502546c.

(27)   Xu, S. Y.; Liu, C.; Alidoust, N.; Neupane, M.; Qian, D.; Belopolski, I.; Denlinger, J. D.;
       Wang, Y. J.; Lin, H.; Wray, L. A.; Landolt, G.; Slomski, B.; Dil, J. H.; Marcinkova, A.;
       Morosan, E.; Gibson, Q.; Sankar, R.; Chou, F. C.; Cava, R. J.; Bansil, A.; Hasan, M. Z.
       Observation of a Topological Crystalline Insulator Phase and Topological Phase
       Transition in $Pb_{1-x}Sn_xTe$. *Nat. Commun.* **2012**, *3* (1), 1192.
       https://doi.org/10.1038/ncomms2191.

(28)   Xiao, Z.; Wang, J.; Liu, X.; Assaf, B. A.; Burghoff, D. Optical-Pump Terahertz-Probe
       Spectroscopy of the Topological Crystalline Insulator $Pb_{1-x}Sn_xSe$ through the Topological
       Phase Transition. *ACS Photonics* **2022**, *9* (3), 765–771.
       https://doi.org/10.1021/acsphotonics.1c01717.

(29)   Du, Z. Z.; Wang, C. M.; Li, S.; Lu, H.-Z.; Xie, X. C. Disorder-Induced Nonlinear Hall
       Effect with Time-Reversal Symmetry. *Nat. Commun.* **2019**, *10* (1), 3047.
       https://doi.org/10.1038/s41467-019-10941-3.

(30)   Nandy, S.; Sodemann, I. Symmetry and Quantum Kinetics of the Nonlinear Hall Effect.
       *Phys. Rev. B* **2019**, *100* (19), 195117. https://doi.org/10.1103/PhysRevB.100.195117.

(31)   Isobe, H.; Xu, S. Y.; Fu, L. High-Frequency Rectification via Chiral Bloch Electrons. *Sci.
       Adv.* **2020**, *6* (13), eaay2497. https://doi.org/10.1126/sciadv.aay2497.

(32)   Xiao, C.; Du, Z. Z.; Niu, Q. Theory of Nonlinear Hall Effects: Modified Semiclassics
       from Quantum Kinetics. *Phys. Rev. B* **2019**, *100* (16), 165422.
       https://doi.org/10.1103/PhysRevB.100.165422.

(33)  König, E. J.; Dzero, M.; Levchenko, A.; Pesin, D. A. Gyrotropic Hall Effect in Berry-
      Curved Materials. *Phys. Rev. B* **2019**, *99* (15), 155404.
      https://doi.org/10.1103/PhysRevB.99.155404.

(34)  He, P.; Isobe, H.; Zhu, D.; Hsu, C.-H.; Fu, L.; Yang, H. Quantum Frequency Doubling in
      the Topological Insulator $Bi_2Se_3$. *Nat. Commun.* **2021**, *12* (1), 698.
      https://doi.org/10.1038/s41467-021-20983-1.

(35)  Itahashi, Y. M.; Ideue, T.; Hoshino, S.; Goto, C.; Namiki, H.; Sasagawa, T.; Iwasa, Y.
      Giant Second Harmonic Transport under Time-Reversal Symmetry in a Trigonal
      Superconductor. *Nat. Commun.* **2022**, *13* (1), 1659. https://doi.org/10.1038/s41467-022-
      29314-4.

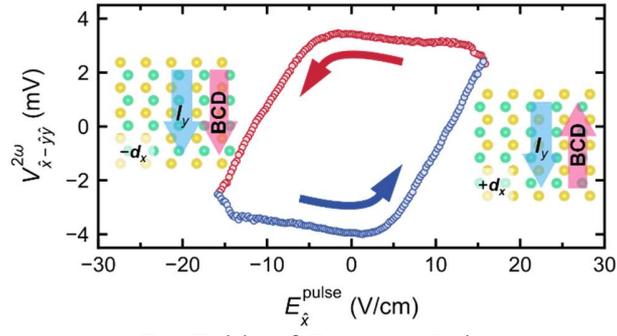

For Table of Contents Only